\def\year{2022}\relax
\documentclass[letterpaper]{article} 
\usepackage{aaai22}  
\usepackage{times}  
\usepackage{helvet}  
\usepackage{courier}  
\usepackage[hyphens]{url}  
\usepackage{graphicx} 
\usepackage{natbib}  
\usepackage{caption} 
\DeclareCaptionStyle{ruled}{labelfont=normalfont,labelsep=colon,strut=off} 
\usepackage{algorithm}
\usepackage{algorithmic}

\usepackage{xcolor}
\newcommand{\answerYes}[1]{\textcolor{blue}{#1}} 
\newcommand{\answerNo}[1]{\textcolor{teal}{#1}} 
\newcommand{\answerNA}[1]{\textcolor{gray}{#1}} 

\usepackage{newfloat}
\usepackage{listings}
\floatstyle{ruled}
\newfloat{listing}{tb}{lst}{}
\floatname{listing}{Listing}
\title{Visibility vs. Engagement: How Two Indian News Websites Reported on LGBTQ+ Individuals and Communities during the Pandemic}
\author {
    Dhruvee Birla,\textsuperscript{\rm 1}
    Nazia Akhtar, \textsuperscript{\rm 1}
}
\affiliations {
    \textsuperscript{\rm 1} International Institute of Information Technology, Hyderabad\\
    dhruvee.birla@research.iiit.ac.in, nazia.akhtar@iiit.ac.in 
}

\begin{document}
\raggedbottom
\maketitle

\begin{abstract}

{In India, online news media outlets were an important source of information for
people with digital access during the COVID--19 pandemic. In India, where “transgender” was legally recognised as a category only in 2014, and same--sex marriages are yet to be legalised, it becomes crucial to
analyse whether and how they reported the lived realities of
vulnerable LGBTQ+ communities during the pandemic. This study analysed
articles from online editions of two English--language newspaper
websites, which differed vastly in their circulation figures---}\textit{The
Times of India}{~and }\textit{The Indian Express}{.}

{The results of our study suggest that these newspaper websites published
articles surrounding various aspects of the lives of LGBTQ+ individuals
with a greater focus on transgender communities. However, they }{lacked
quality and depth}. Focusing on the period spanning March 2020 to
August 2021, we analysed articles {using sentiment analysis and topic modelling.} We also compared our results to the period before the pandemic (January 2019 -- December 2019) to understand the shift in topics, sentiments, and stances across the two newspaper websites. A manual analysis of the articles indicated that the language used in certain articles by \textit{The Times of India}{~was transphobic and obsolete. Our study captures the visibility and representation of the LGBTQ+ communities in Indian newspaper websites during the pandemic.}

\end{abstract}

\section{Background}
{Coronavirus (COVID-19) was declared a pandemic on 11 March 2020. A highly infectious disease caused by the acute respiratory syndrome
coronavirus 2 (SARS-CoV-2), it was first detected in China in December
2019 and has since rapidly spread to other countries. From December 2019 to June 2023, over 13 million vaccines were
administered, and by August 2023, over 760 million cases and 6.9 million
deaths were recorded. In
addition to this}{, 67\% of the global population by November 2023 had
been vaccinated with a complete primary series of a COVID-19 vaccine,
which in most countries comprised of two doses \citep{who23}. }

{}

{With the onset of the pandemic in 2020, regulations to check the spread
of the disease through physical isolation were implemented in different
countries. These checks/measures with the goal of social distancing were
put in place to different degrees in different regions. While some regions implemented measures
by choosing total isolation, others implemented minimal social
distancing regulations \cite{hiscott2020global}. }

{}

{Previous widespread outbreaks of infectious diseases have brought in
their wake an intensification of gender inequalities in access to
healthcare, social support, education, and employment at a global level
\cite{wenham2020women, stemple2016human}. The COVID--19 pandemic was no
exception to this norm \cite{al2021investigating, gausman2020sex, phillips2020addressing, yuan2023minority, carli2020women, alon2020impact, flor2022quantifying, fish2021sexual}{. \citet{adamson2022experiences} performed a
global cross-sectional analysis---involving 79 countries---to
characterise the degree to which the levels of violence and
discrimination against vulnerable communities have changed amid
SARS-CoV-2. It was found that ethnic minorities, disabled
people, and those who identified as gay or queer experienced more
discrimination from government representatives, state apparatuses, and
healthcare providers \cite{adamson2022experiences}. Particularly in the context
of gender and sexual minorities, LGBTQ+ individuals and groups living in
different countries also experienced an intensification of
discrimination, prejudice, and violence during the COVID--19 pandemic,
leading to unique experiences during the pandemic \cite{ganguly2021transgender, roy, konnoth2020supporting, lucas2022lgbtq+, langness, whittington2020lives}. To make matters worse,
laws that added to the stigma against LGBTQ+ individuals were passed
and/or maintained in many countries, including Singapore and the United
States.}

{}

{Online news outlets were a prominent source of information during the
COVID--19 pandemic for those who had digital access, covering various
topics, from the spread of the virus to government regulations,
measures, and updates to constrain the virus \cite{ghasiya2021investigating}. According to a survey
conducted in 2020 by KPMG India Analysis, news consumption through
digital applications increased, thus decreasing traditional
news consumption by approximately 38--40\%\cite{kpmg2020media}. 
The 2021 Digital News Report by the Reuters Institute for the Study of
Journalism shows that 82\% of
Indians use online sources to read news, an increase from 56\% in 2019
\cite{aneez2019india, newman2021reuters}.\footnote{Indian national print media suffered from a terrible recession during the COVID--19
pandemic. Disruptions in newspaper distribution due to lockdowns, high
production costs, financial crisis due to a decline in advertisements
revenue, and cost--cutting targets by laying off journalists and other
employees are a few difficulties faced by the Indian print media \cite{saxena2021impact}. Despite these challenges, during the COVID--19 pandemic, Indian
news media played a major role in bringing pandemic--related information
to their audience. }

{Mainstream media outlets have the power to decide what news the
readership is informed of and how that information is presented \cite{mcquail1987mass}. Given the popularity of online news media during the pandemic,
online news media has played a crucial role in the spread and narration
of information as well. In the past, specifically in the context of the
transgender community, research conducted exposes news media outlets of
their marginalised and stereotypical narratives and representations of
trans people \cite{barker2013fixing, capuzza2016improvements, gupta2019response, hackl2016chelsea, aakerlund2019representations}. The language, or narration, used in news media outlets is used to construct issues in a certain way. It
}{holds political power to either maintain or deconstruct
ideologies based on beliefs rooted in conventions constituting hegemonic
power relations \cite{fairclough2013language, willox2003branding}. Therefore, the topics
chosen by news media outlets to inform their audience and the
articles\textquotesingle{} narration are crucial in forming public
opinion.}

{}

{To date, much research has been conducted in different geographical
contexts to understand what news was conveyed by the online news media
during the COVID--19 pandemic \cite{bogovic2021topic, liu2020health, wan2021topic, ghasiya2021investigating}. Qualitative research has
been conducted using critical discourse analysis of the
news media articles \cite{wiktorek2015prohibit, hindarto2022investigating} along with a
thematic analysis of media framing \cite{morrison2021newsworthy}. \citet{jain2021analysis} created a hybrid model that can analyse the overall effect of digital news content in India through a hybrid approach of sentiment analysis to classify all headlines. Particularly in the context of LGBTQ+ communities, a study by \citet{aakerlund2019representations} analysed the representation of transgender people in approximately 16000 Swedish newspaper articles published in the period 2000--2017 using topic modelling and critical discourse analysis.}{~The results indicated that the media positioned transgender people by trivialising trans expressions, incorporating these expressions into the gender binary and excluding them by framing them as ``deviant" \cite{aakerlund2019representations}.
Additionally, several studies and articles compiled in} \citet{arora2023media} {investigated Asian media narratives during the
COVID--19 pandemic. It brought to light how the COVID--19 pandemic
affected different nations differently and intensified
discrimination against communities marginalised based on gender, caste,
class, nationality and religion. }

{}

{Studies have also found that n}{ews media still lack fair
representation}{~of LGBTQ+ communities and their activities, such as the
Pride festival \cite{morrison2021newsworthy, semykina2018media}. While online news
articles covered issues relating to the LGBTQ+ communities, they made
little effort to provide a comprehensive picture and vital context. The
lack of content can be found in the lack of paragraphs and small lengths
of the articles \cite{semykina2018media, listiorini2022news}. Research
conducted by \citet{raj} investigates the missing gendered narratives in
Indian news media and the impact of COVID--19 on women. It analysed
articles published by the top three English dailies, namely, }\textit{The Times
of India}{,}\textit{~The Hindu}{,}{~}{and}\textit{~Hindustan Times}{, from 25 March
2020 to 25 June 2020. It found that while these national dailies covered
the difficulties faced by healthcare workers, they excluded women
healthcare providers and did not mention the difficulties faced by
the LGBTQIA+ communities during the pandemic \cite{raj}. {Therefore, the coverage of Indian news media
remained largely gender--blind, rendering the gendered
dimensions of the pandemic invisible \cite{raj}. }

{}

{Our study focuses on narratives employed by online Indian news media
and their coverage of the lived realities of LGBTQ+ communities.}

{}

{~We aim to answer the following questions: }

{}

\begin{enumerate}
\item
  {What was the thematic scope of coverage of LGBTQ+ communities during
  the COVID-19 pandemic in the online version of }\textit{Times of India}{, the
  newspaper with the highest circulation figures in India? In contrast,
  what was this coverage like in }\textit{The Indian Express}{, which has a
  relatively smaller circulation?}
\item
  {What sentiment and stance can be detected in the articles using state-of-the-art methods? How and why are they similar or different from the stance and
  sentiment expressed before the pandemic?}
\item
  {What were the narratives and attitudes in the two newspapers in the
  context of LGBTQ+ communities, discernable from the identified themes
  (topics) and sentiment? }
\end{enumerate}

{}

{Our objectives, therefore, are as follows: }

{}

\begin{enumerate}
\item
  {To document, collect, and analyse themes of online articles published
  by }\textit{The Times of India}{~and }\textit{The Indian Express}{~from the pandemic
  period;}
\item
  {To detect, classify, and interpret the sentiment and stance of the
  articles posted online during the pandemic period by }\textit{The Times of
  India}{~and }\textit{The Indian Express}{, and compare them}{~with the same
  before the pandemic; and }
\item
  {By synthesising insights from topics, sentiments, and stances, to
  arrive at insights about the narratives followed by two of the highest
  circulating Indian newspapers while reporting on LGBTQ+ communities
  during the pandemic period.}
\end{enumerate}

\section{Data}
{For this study, we chose }\textit{The Times of India}{~(}{TOI}{) and }\textit{The
Indian Express }{(}{IE}{)}{, which are English--language dailies. }{We considered circulation figures while deciding on the news outlets for this research.}{~}\footnote{~We ran into
practical issues with other English--language dailies that fulfil the
criteria but were not considered due to restrictions in accessing their
online archives.}

{}

{Both }\textit{TOI}{~and }\textit{IE}{, in their own individual way, had a sustained
engagement with LGBTQ+ issues even before the COVID--19 pandemic.}

{According to 2023 statistics, \textit{TOI} has
approximately 16 million readers across 36 cities in India, followed by
\textit{The Hindustan Times} \cite{audit}. }\textit{TOI}{~has
been a visible supporter of ~LGBTQ+ communities, evidenced by a campaign
they started, \#TimesOutAndProud, in May 2019.
This initiative seeks to make ``daily life a little easier for the
LGBTQIA+ community" and raise awareness about their lived experiences.
As part of this campaign, a series of initiatives were taken across
digital platforms, television, radio and on-ground activities, which
included the creation of digital films and organisation of pride
parades, among other activities \cite{times}. }

{}

\textit{The Indian Express}{, a popular English--language daily, is a }{lesser
circulating newspaper}{~with nearly 1.6 million readers according to the
Indian Readership Survey 2017 \cite{irs2019} and is not present in the list of highest
circulating English--language newspapers published by the Audit Bureau
of Circulations. Even though }\textit{IE}{~does not have a campaign or other
activities providing a platform for the LGBTQ+ communities, they harness
the power of written expression to raise awareness about LGBTQ+ issues.
This newspaper website has informed its audience about the difficulties
with the process of coming out, and the atrocities
faced by LGBTQ+ individuals worldwide alongside other issues for years before the
pandemic.}\footnote{Examples of
\textit{IE} articles that cover these issues are titled, "What they didn't tell
you: Coming out is a never-ending process", "Turkey bans all LGBTI
events in Ankara, citing security", and "How
to start a conversation on LGBTQI issues at home? Keshav Suri has
answers"}


{}

These two particular newspaper websites were chosen to examine whether
there is a contrast between the quality, quantity, and attitudes
pertaining to LGBTQ+ communities prevalent across newspapers with
differing circulation numbers during the pandemic period. 


{Using the archives and sitemap of these two channels, we gathered data
from two timelines---March 2020 to August 2021 and January 2019 to
December
2019.}\footnote{~All articles
were ethically scraped and were in accordance to the robots.txt file of
the news outlet.}{~The
former timeline was explicitly chosen as the pandemic period in our
study since the World Health Organisation (WHO) declared COVID--19 a
pandemic in March 2020, and by August 2021, most people with access had
received their first vaccination shot, with many high--income countries
offering booster vaccines to their residents \cite{world2022director}. }

{}

{News articles gathered from these time periods were filtered using
keywords such as transgender and
LGBT.}\footnote{~The keywords
used were, "Transgender", "Trans", "Lesbian", "Gay", "LGBT", "LGBT+",
"LGBTQ", "LGBTQ+", "LGBTQIA", "LGBTQIA+", and "Queer".}{~We collected }{1576 articles} {(477 from }\textit{TOI}{~and }{1099 from }\textit{IE}{) during the pandemic
period and 1461 articles (659 from }\textit{TOI}{~and 790 from }\textit{IE}{) from
before the pandemic period.}{~We analysed two primary components of the
articles---headlines and article content. We conducted sentiment
analysis on both the components using distil--RoBERTa--base and ChatGPT--3.5. To complement the results from sentiment
analysis, we performed stance detection using the TESTED model trained
on the Fake News Challenge--1 (FNC--1) dataset \cite{pomerleau2017fake}.\footnote{Codebase: {\textcolor[HTML]{000099} {https://github.com/copenlu/TESTED/tree/main}}} Furthermore, we performed topic modelling on the
articles\textquotesingle{} content using BERTopic.}


\section{Methods}

\subsection{Topic Modelling}

{}

{We use BERTopic to perform topic modelling on the
articles\textquotesingle{} content. BERTopic uses a three--step process
to generate topics. Firstly, the documents are converted into their
embedding representation, for which we used the }\textit{all-MiniLM-L6-v2 }{model.
Following this, the embeddings\textquotesingle{} dimensionality is
reduced, which we performed using the Uniform Manifold Approximation and Production (UMAP) method. The last step is clustering, which was performed using
HDBSCAN. The last step involves extracting topic
representations from the document clusters using a class-based TF--IDF
variant \cite{grootendorst2022bertopic}. 


{}

\subsection{Sentiment Analysis}

{}

We performed sentiment analysis on news headlines and the content of the article. We analysed headlines since they are the bridge that connects the reader with the
article \cite{palau2016reference}. With the age of digitisation, the way news
headlines are written has also evolved. Headlines are structured at the
expense of quality to make them more attractive to users and secure high
positions in search results \cite{chakraborty2016stop, scacco2016investigating}. Due to the abundance of online resources, consuming content
online has become hurried and superficial \cite{jiang2019prompts, chakraborty2016stop}. Most people review and determine the context
of the news article by scanning through the headline instead of reading
the article\textquotesingle s content \cite{rieis2015breaking}. Therefore,
analysing news headlines becomes crucial when many readers get their
information solely from them.}{~}{Furthermore, we perform sentiment analysis on the articles' content to determine if the headline's sentiment accurately represents that of the article. 

{}

{RoBERTa is a robustly }{optimised}{~BERT approach that has achieved
SOTA results in RACE, GLUE, and SQuAD \cite{liu2019roberta}. Sentiment analysis on news headlines and articles was performed using distil--RoBERTa--base, which is pretrained for sentiment
analysis on financial news headlines.\footnote{This model is hosted on HuggingFace by Manuel Romero. Link: {\textcolor[HTML]{000099} {https://shorturl.at/syGSV}}} We also used ChatGPT--3.5, which has
been gaining ground rapidly. Within two months of its launch, 100
million users were using the AI bot, with 13 million unique visitors per
day in January 2023 \cite{hu}. In the task of sentiment analysis, it
was found that ChatGPT's zero--shot capabilities is on par with
finetuned BERT. In contrast, with the few--shot
prompting technique, ChatGPT\textquotesingle s performance increases
significantly, surpassing finetuned BERT in certain domains \cite{wang2023chatgpt, qin2023chatgpt}. In this research, we will be using RoBERTa and ChatGPT–3.5 for the task of sentiment analysis in the context of LGBTQ+ communities. This approach is motivated by the relatively limited exploration of ChatGPT's performance in analyzing sentiments specifically related to queer identities. By comparing RoBERTa and ChatGPT, we aim to evaluate their effectiveness and nuances in understanding and processing LGBTQ+ narratives.
}

{}

{The reason behind using these two models is to see how well }{a
maverick entrant like}{~ChatGPT performs against a finetuned post--BERT model in the domain of LGBTQ+--related news.}

{Previous research has stressed the importance of high--quality datasets
for research on sensitive topics concerning vulnerable and marginalised
groups, which requires all actors working on the research, including
annotators, to be aware of the lived realities of the communities being
studied \cite{kumaresan2023homophobia}. Therefore, three LGBTQ+
annotators labelled 1576 headlines as positive, negative, or neutral. To measure the reliability (level of
agreement between the raters) of the annotated dataset, we used
Krippendorff\textquotesingle s Alpha, which measures the reliability
between an arbitrary number of raters. The inter-rater score was 0.58
using Krippendorff\textquotesingle s Alpha. {The labelled dataset was divided into 1000, 300, and 276 headlines for training, validation, and testing, respectively.} The majority sentiment for
each headline from this labelled dataset was used to finetune the pretrained
distil--RoBERTa--base model, and few were randomly selected as examples to prompt ChatGPT–3.5 \cite{NEURIPS2020_1457c0d6}.}

{}

While performing sentiment analysis on the articles' content using ChatGPT, generating results took longer due to the large input size, and ChatGPT had to constantly be reminded of either the task or the output format (see Appendix \ref{sec:appendixGPT}). Therefore, we performed sentiment analysis on the content using
only a finetuned distil-RoBERTa-base.


\subsection{Stance Classification}

{Stance }{detection (also known as stance classification and stance
prediction) is used to determine the attitude/position of the author
towards a target of interest \cite{mohammad2016semeval, kuccuk2020stance}.
This is an important tool for assessing the stances towards a particular
target and can be used in various tasks, including but not limited to
identifying the leanings of media outlets \cite{stefanov2020predicting}. }

{}

{The TESTED framework proposed by \citet{arakelyan2023topic} can predict
stances across various domains and has been evaluated on a multi--domain
dataset consisting of 16 individual datasets, which are grouped into 4
different categories--News, Debates, Social Media, and Various \cite{hardalov2021cross}.}{~A
topic--guided diversity sampling method has been introduced within
TESTED, and the generated multi--domain supervised training sets were
used to create a stance detection model by finetuning a pre-trained
language model, }\textit{roberta--large}{, using a contrastive objective. This
framework has achieved SOTA results on 10 out of the 16 datasets, where
it obtained an F1--score of 83.17 on in--domain experiments and 72.51 on
out--of--domain experiments using the FNC--1 dataset. To pre--train TESTED, we used the FNC--1
dataset, which is a labelled dataset that contains the stance of news articles towards their corresponding headlines. The stances could be one of the four
labels---agree, disagree, discuss, or unrelated. }

\section{Results}

\subsection{Topic Modelling}

{}

{Topic modelling results indicate that during the pandemic, the top topic in }\textit{TOI}{~is
Pride Celebrations and representation in popular culture, and in }\textit{IE}{,
it}{~is Pride celebrations. Tables \ref{table1} and \ref{table2} mention
some of the interpreted topics alongside the top 5 keywords appearing in the topic. Using data from the pandemic period, the model achieved a coherence score of 0.77 using }\textit{IE} data{~and 0.79 using }\textit{TOI} data.

\begin{table}[htbp]
\resizebox{.95\columnwidth}{!}{
\begin{tabular}{|p{3cm}|p{6cm}|}
\hline
\textbf{Topic} & \textbf{Interpreted Topic} \\
\hline
Leave, employees, India, gender, and women & Diverse and gender-inclusive policies in industries. \\
\hline
Marriage, sex, court, family, and freedom & Exploitative jobs, right to privacy and forced proximity of abusive family members. \\
\hline
Voters, polling, candidates, election, and district & American politics \\
\hline
Covid, cases, said, vaccination and 19 & Disruption of normalcy, vaccination centres, medical camps and other support services \\
\hline
Police, arrested, accused, transgender, and woman & Legal cases that involved LGBTQ+ individuals \\
\hline
Transgender, help, community, people, and government & Welfare programs and policies implemented for transgender people \\
\hline
Film, people, pride, one, and community & Pride Celebrations and representation in popular culture \\
\hline
\end{tabular}
}
\caption{Interpreted topics of \textit{TOI} from the pandemic period}
\label{table1}
\end{table}

\begin{table}[htbp]
\resizebox{.95\columnwidth}{!}{
\begin{tabular}{|p{3cm}|p{6cm}|}
\hline
\textbf{Topic} & \textbf{Interpreted Topic} \\
\hline
Games, Olympics, athletes, Tokyo, and Hubbard & Global LGBTQ+ representation in sports \\
\hline
Transgender, people, also, community, and government & Welfare programs and policies implemented for transgender people \\
\hline
Biden, Trump, President, first, and house & American politics \\
\hline
Pride, community, month, pride month, and rainbow & Pride Celebrations \\
\hline
People, brand, fashion, year, and one & Social movements \\
\hline
Page, actor, Elliot, like, and show & LGBTQ+ representation in popular culture \\
\hline
Film, also, show, one, and actor & Popular culture \\
\hline
Poland, law, lgbt, Turkey, and European & Global politics \\
\hline
People, book, Rowling, trans, and women & Discussions around the representation of LGBTQ+ communities in popular culture \\
\hline
\end{tabular}
}
\caption{Interpreted topics of \textit{IE} from the pandemic period}
\label{table2}
\end{table}




{}




{}

{In contrast, the most popular topic covered by }\textit{TOI}{~before the
pandemic is \textquotesingle Legal cases that involved LGBTQ+
individuals\textquotesingle, whereas by }\textit{IE}{~is
\textquotesingle LGBTQ+ Representation in popular
culture\textquotesingle.}{ Using data from before the pandemic period, the model achieved a coherence score of 0.68 using
}\textit{IE} data{~and 0.82 using }\textit{TOI} data{. }

{}

\subsection{Sentiment Analysis}

{}

{Before fine–tuning, the pre–trained distil--RoBERTa--base achieved an accuracy of
51.6\%, with a precision of 50, recall of 38 and f1-score of 36. Whereas, ChatGPT obtained an accuracy of 58.8\%, with a precision of
68, recall of 70 and f1-score of 60.} This indicates that zero--shot performance of ChatGPT is better than domain--specific pretrained distil--RoBERTa--base in the domain of LGBTQ+--related news.

{}

{The distil--RoBERTa--base model was finetuned using five--fold
cross--validation. }{After finetuning, the model achieved an accuracy of 77.5\%, with a precision of 81.0, recall of 77.5
and f1-score of 77.7. On the other hand, few-shot ChatGPT obtained an accuracy of
75\%, with a precision of 76, recall of 75 and f1-score of 75 (see Appendix \ref{sec:appendixCM} for class--wise metrics)}. This indicates that the few--shot technique considerably increased the performance of ChatGPT. However, in the domain of LGBTQ+--related news, it is inferior than finetuned distil--RoBERTa--base model. 

{}




{}




{}

{During the pandemic period,}{ both distil--RoBERTa--base and ChatGPT–3.5,
to varying degrees, depicted that }\textit{TOI}{~posted positively--toned
headlines the most whereas,
most headlines were neutral--toned by }\textit{IE} (see Tables \ref{table8} and \ref{table9}). In contrast, before the pandemic, distil--RoBERTa--base depicted that most of the headlines by both \textit{TOI}{~and }\textit{IE}{~were
neutral--toned (see Tables \ref{table10} and \ref{table11}). 

\begin{table}[htbp]
\centering
\resizebox{.95\columnwidth}{!}{
\begin{tabular}{|c|c|c|}
\hline
\textbf{Sentiment} & \textbf{TOI} & \textbf{IE} \\
 \textbf{(in \%)} & \textbf{[during pandemic]} & \textbf{[during pandemic]}\\
\hline
Positive & 37.3 & 26.3 \\
Negative & 33.8 & 31.8 \\
Neutral & 28.9 & 41.9 \\
\hline
\end{tabular}
}
\caption{The polarity of news headlines assigned by distil–RoBERTa–base}
\label{table8}
\end{table}

{}

\begin{table}[htbp]
\centering
\resizebox{.95\columnwidth}{!}{
\begin{tabular}{|c|c|c|c|c|}
\hline
\textbf{Sentiment} & \textbf{TOI} & \textbf{IE} \\
 \textbf{(in \%)} & \textbf{[during pandemic]} & \textbf{[during pandemic]}\\
\hline
Positive & 45.9 & 23.7 \\
Negative & 23.8 & 15.1 \\
Neutral & 30.1 & 61.0\\
\hline
\end{tabular}
}
\caption{The polarity of news headlines assigned by ChatGPT}
\label{table9}
\end{table}

\begin{table}[htbp]
\centering
\resizebox{.95\columnwidth}{!}{
\begin{tabular}{|c|c|c|}
\hline
\textbf{Sentiment} & \textbf{TOI} & \textbf{IE} \\
 \textbf{(in \%)} & \textbf{[before pandemic]} & \textbf{[before pandemic]} \\
\hline
Positive & 35.8 & 19.3 \\
Negative & 27.5 & 26.8 \\
Neutral & 36.7 & 53.9 \\
\hline
\end{tabular}
}
\caption{The polarity of news headlines assigned by distil–RoBERTa–base}
\label{table10}
\end{table}

\begin{table}[htpb]
\centering
\resizebox{.95\columnwidth}{!}{
\begin{tabular}{|c|c|c|}
\hline
\textbf{Sentiment} & \textbf{TOI} & \textbf{IE} \\
 \textbf{(in \%)} & \textbf{[before pandemic]} & \textbf{[before pandemic]} \\
\hline
Positive & 48.1 & 34.2 \\
Negative & 23.1 & 24.6 \\
Neutral & 28.8 & 41.2 \\
\hline
\end{tabular}
}
\caption{The polarity of news headlines assigned by ChatGPT}
\label{table11}
\end{table}

Sentiment analysis on the content of the articles using distil--RoBERTa--base indicates \textit{TOI} wrote positive--toned articles and \textit{IE} wrote neutral--toned articles the most both during and before the pandemic period (see Appendix~\ref{sec:appendixSA}).}




{}

{}

\subsection{Stance Classification }

{}

{The result from stance classification indicates that almost all of the
headlines published by }\textit{TOI}{~and }\textit{IE} both before and during the pandemic period {accurately represent the
article\textquotesingle s content. However, most of the remaining
articles have content unrelated to the headline (see Tables \ref{table15} and \ref{table16}). The percentage of \textquotesingle agree\textquotesingle{}
stance increased during the pandemic in }\textit{IE}{~articles, whereas it
dropped in }\textit{TOI}{~articles.}

\begin{table}[htbp]
\centering  
\resizebox{.95\columnwidth}{!}{
\begin{tabular}{|l|c|c|}
\hline
\textbf{Stance} & \textbf{TOI} & \textbf{TOI}\\
 \textbf{(in \%)} & \textbf{[during pandemic]} & \textbf{[before pandemic]}\\
\hline
Agree & 85.53 & 87.78\\
\hline
Disagree & 1.05 & 0.30 \\
\hline
Discuss & 0.84 & 1.34\\
\hline
Unrelated & 12.58 & 10.58\\
\hline
\end{tabular}
}
\caption{Stances of \textit{TOI} articles}
\label{table15}
\end{table}

\begin{table}[htbp]
\centering  
\resizebox{.95\columnwidth}{!}{
\begin{tabular}{|l|c|c|}
\hline
\textbf{Stance} &\textbf{IE} & \textbf{IE}\\
 \textbf{(in \%)} & \textbf{[during pandemic]} & \textbf{[before pandemic]}\\
\hline
Agree  & 86.57 & 82.93 \\
\hline
Disagree & 1.19 & 0.89 \\
\hline
Discuss & 1.92 & 2.42 \\
\hline
Unrelated & 10.32 & 13.76 \\
\hline
\end{tabular}
}
\caption{Stances of \textit{IE} articles}
\label{table16}
\end{table}

\section{Discussion}

{During the pandemic period, the number of articles retrieved from }\textit{IE}{~covering
}{LGBTQ+}{--related events was more than twice than those recovered
from }\textit{TOI}{. Moreover, the average length
of the articles retrieved from }\textit{TOI} (2563 characters) was considerably lesser than that of \textit{IE} (3909 characters).
Prior research has found that while the online news articles covered
issues relating to the LGBTQ+ communities, they made little effort to provide substantial and meaningful coverage of the lived realities of LGBTQ+ individuals, thus throwing into question the
credibility and reliability of the article. Research has found the lack
of content to consist of the lack of paragraphs and small wordcount
\cite{semykina2018media, listiorini2022news}, as can also be noticed
in the articles we retrieved. }

{}

{The number of articles that write about issues that do not address the urgent realities of the pandemic overshadow the articles written on the lived
realities of LGBTQ+ individuals during the time of the pandemic. }\textit{IE}{~offers a
valuable qualitative contrast to the coverage }\textit{TOI}{~has. }\textit{TOI}{~has
taken up many initiatives to voice the range of difficulties faced by
LGBTQ+ individuals and communities. However, when we start looking
closely at its material and compare it to }\textit{IE}{, we find that there is
a significant difference in the quality and meaningfulness of the
articles between the two newspaper websites. The number of articles
written about other important aspects of their lives, such as
discrimination by governme}{nt representatives, le}{gislative
institutions, violence and harassment by family members and other
patriarchal institutions, had limited to no coverage. This coverage was
also mostly focused on the transgender community. These themes and
patterns observed in our methods---topic modelling, sentiment analysis,
and stance detection---have been discussed in detail in this section. Furthermore, we have qualitatively analysed a part of our dataset to arrive at a nuanced and granular understanding of our quantitative results.}

{}

\subsection{Topic Modelling}

{The major topics covered by }\textit{TOI}{~and }\textit{IE}{~during the pandemic were
representation in popular culture, Pride celebrations, legal cases
involving the LGBTQ+ communities surrounding a wide range of issues,
welfare programs during the COVID--19 pandemic, disruptions in
professional as well as personal spheres, and vaccination--related
events. Articles on Pride Celebrations and Popular Culture have
predominated before and during the pandemic. 

Through our manual analysis of the dataset, we found that {even though }\textit{TOI}{~provides greater visibility to the LGBTQ+
communities in Indian society, their writing and the language used
continue to be problematic. The purpose of news media outlets is to
inform, and there is an expected corresponding commitment to authentic
and responsible reporting. This, however, was not fully realised in the
}\textit{TOI}{~reporting of LGBTQ+ communities. 
}

{}

{An example}{ of this is the usage of the pejorative ter}{m ``eunuch" to
}{date.}\footnote{~They
mention the term in the article titled, "Eunuch kills lover of over a
decade over domestic dispute."}{~This term is considered dehumanising and objectifying, and is linked to other pejorative words, such
as ~``effeminate,'' ``emasculated,'' and ``impotent.'' In addition to
this, }\textit{TOI}{\textquotesingle s articles have been found describing
transgender as a "condition".}\footnote{~Please see
the article titled "Kanpur: Man files FIR against in-laws for misleading
him to marry transgender."}{~}{The
poor--quality writing could be due to submission by stringers, which
have further not been verified by the editors of the media outlet,
indicating the lack of meaningful interest in LGBTQ+ events. Along with
this, the reporting in some articles seems to be superficial and lacks
quality. Previous research has also found that often, through the use of
offensive language, such as the derogatory terms used in \textit{TOI} articles,
trans people are framed as deceptive \cite{capuzza2016improvements, mackenzie2009media, aakerlund2019representations}}.

{}

Even though \textit{IE}{~does not write extensively about LGBTQ+ communities,
it still manages to write a few articles on LGBTQ+--centric experiences
during the COVID--19 pandemic, as observed from the results of topic
modelling, such as experiences in schools, HIV/AIDS, loneliness, lack of
financial resources and forced proximity to homophobic family members.
Instead of simply stating facts and conveying information, the newspaper
does more to educate society about the different experiences of
different vulnerable and marginalised communities, including LGBTQ+
communities, across the country. Articles titled ``Life in the time of
social distancing: Confined in homes, people battle restlessness,
anxiety", ``Home and unsafe", and ``Hit hard by lockdown: Transgender
community stands isolated with no financial resources" are a few such
examples.}

{}

{Some overlapping topics covered by }\textit{IE}{~and }\textit{TOI}{~that can be
observed in the results of our topic model---despite varying extents and
quality of coverage---are sex work, Pride Celebrations, HIV/AIDS, gender
dysphoria, popular culture, and welfare programs. As interpreted from
the keywords, the welfare programs discussed are initiatives taken up by
governments, NGOs, and groups of people to provide free kits, such as
masks and ration supplies, to the transgender communities{. }

Additionally, as depicted by the topic modelling results, the articles touch upon HIV/AIDS, noting the demand of transgender people for special packages that include medications for hormone therapy. During the HIV epidemic, stigma surrounded the LGBTQ+ communities, and the
pandemic brought back memories of the discrimination. This
discrimination prevailed in many aspects of LGBTQ+
people\textquotesingle s lives, such as health care. Additionally, sex
reassignment surgeries (SRS) could not be performed during the pandemic,
which led to further complications such as hormonal imbalance \cite{stevens2021natural}. }

{}

{The reason for covering sex work, another topic that appeared in the topic modelling results, is multi--dimensional. It is
established that LGBTQ+ individuals face reduced opportunities of being
accepted in mainstream jobs. This was also highlighted in the articles
published by }\textit{TOI}{~and }\textit{IE}{, which reported this issue several times
during the pandemic. Sex work, dancing at weddings and festivals, and
seeking alms could not continue to be a source of income during the
pandemic due to regulations such as social distancing, which led to
economic instability.} Unfortunately, no other means of livelihood
and a lack of officially recognised identity cards such as Aadhar cards
and ration cards created obstacles for the transgender communities
during the
pandemic.\footnote{~The lack
of official documents is often due to the incorrect gender assigned
during their birth, which later changed through transition.}


{Pride Celebrations were an event that many looked forward to during the
pandemic as a form of social and communal action. These celebrations
also usually raise funds for various LGBTQ+--related activities and support services. However, due to
the pandemic, these celebrations could not take place or were conducted
online, which led to insufficient funds \cite{konnoth2020supporting}.
The lack of a support group further intensified feelings of
depression and anxiety, which is also noticeable from articles published
by }\textit{TOI}{~and }\textit{IE}{. These articles report the starting of special
helplines for women and LGBTQ+ communities and share that these helplines
received calls from young children and teenagers forced to live with
abusive families, isolated and with no support network, susceptible to
self--harm during the
pandemic.}

{Through the topic modelling results, we also found a few articles written about global politics related to
the LGBTQ+ community. These topics are alive in the public discourse,
often of interest to the Indian readership due to ongoing legislative
and litigation procedures. These articles were filtered into our dataset
due to the conversations around the rights of LGBTQ+ communities and
because, }{during the pandemic, many countries upheld or modified laws
that added to the already existing stigma against the communities}{.
Specifically in the American context, during the COVID--19 pandemic
(June 2020), the Trump administration (2017--2021) reversed the
protections put in place by the Obama administration (2009--2017). Reversing these protections allowed
healthcare practitioners to deny necessary services to transgender
people during the COVID--19 pandemic. This led to added stress and
anxiety since many required medications, such as ART drugs and hormone
supplements, along with necessary medical care when infected by the
virus. }


{In LGBTQ+ political spaces, another significant issue raised is
representation. Greater representation is vital to increasing acceptance
and normalisation for LGBTQ+ community members in the coming generations
\cite{shah2015no}. Representation in popular culture was the top topic
in }\textit{TOI}{~during the pandemic but was a less significant topic in
}\textit{IE}. These articles are about movies, books, and poems written about
LGBTQ+ individuals, along with festivals organised specifically for
recognising LGBTQ+ popular culture. 

{We came across another topic, Family and Friends, that was of interest
to us because of the centrality of family and institutions such as
schools in an individual's life, but despite its relevance to the
pandemic experiences of LGBTQ+ individuals, there are very few articles
dedicated to this topic.\footnote{Any public space, such
as school, can be a safe place or a place of extreme suffering. The interactions at home and school can lead to gender dysphoria,
which continues to affect them later in life \cite{shah2015no}.}

{}

\subsection{Sentiment Analysis \& Stance Classification}

{}

{A study conducted by \citet{rieis2015breaking} found that headlines with a negative and a positive
polarity garnered higher interest than an article with a neutral--toned
headline \cite{rieis2015breaking}. According to distil--RoBERTa--base results, positive and negative
headlines from both \textit{TOI} and \textit{IE} increased slightly during the pandemic. However, ChatGPT–3.5 results indicate that these headlines
either slightly decreased or remained the same{. }

{}

{A positive sentiment could imply a more positive representation of
LGBTQ+ communities in online media discourse. Our manual analysis
indicated that a major proportion of positively--toned headlines informed of initiatives by the government, NGOs, the general public, and LGBTQ+ individuals to provide
support to the LGBTQ+ communities or individuals in need. On the other hand, headlines with a negative tone
contained information about the difficulties faced by the LGBTQ+
communities as well as crimes committed by LGBTQ+ individuals.
}{However, as mentioned in our discussion on topic modelling results,
the topics covered by these newspaper websites were superficial. For
instance, anti--queer remarks made by famous artists and reported by
}\textit{TOI}{~were also rightly marked negatively, but did not contribute
towards communicating the experiences of the communities, particularly
during the pandemic. }

{}

{No major shift was noticed in the stance and sentiment of the headline and the article's content published by
}\textit{TOI}{~and }\textit{IE}{~from before to during the pandemic. The results also }{indicate that most of the headlines{~accurately represented the sentiment and content of the article.}


\section{Conclusion}

{We found that even though the articles cover various aspects of
the lives of LGBTQ+ individuals, they still fell short in certain ways. \textit{TOI} and \textit{IE} focused more on transgender communities compared to other LGBQ+
communities. { Furthermore}{, }\textit{TOI}{, when compared
to }\textit{IE}{, writes superficial articles on the LGBTQ+ communities, which,
however, does not mean extensive coverage by }\textit{IE}{. This is
noticeable from our topic modelling results that indicate the gaps in covering various aspects of the lives of
LGBTQ+ individuals, the number of articles published, and the brief articles written on their lived
realities. We conclude that the English--language Indian newspaper
website, }\textit{TOI}, does not provide a substantive platform for LGBTQ+
communities to voice their discriminatory experiences during the
COVID--19 pandemic in India.

{Additionally, we found that in the task of news sentiment analysis in the context of LGBTQ+ communities, few--shot ChatGPT is inferior to finetuned domain--specific distil--RoBERTa--base model. }


{}

{Our study provides an overview of the shift in the narratives of
}\textit{TOI}{~and }\textit{IE}{~from before the pandemic to during the pandemic.
Firstly, we have observed a positive development in both }\textit{TOI}{~and
}\textit{IE}{, which covered a wider range of gender--specific issues during
the pandemic, such as gender dysphoria and the effect of private spheres
on the lives of LGBTQ+ communities. However, during a time of crisis, news forums
should put extra effort into conveying news that is fair, unbiased and
gender--sensitive to adequately represent the inequalities present in
society \cite{tshuma2022media}.
The increase in
awareness in both media outlets, however, does not meaningfully capture
the lived realities of LGBTQ+ communities. The gaps in coverage and use
of offensive language render LGBTQ+ communities invisible. An
overwhelming coverage of shallower issues with a minimal emphasis on the
impact of COVID--19 on the lives of LGBTQ+ communities indicates a lack
of interest of the Indian media in covering LGBTQ+ issues. }

\section{Limitations}

{Our research includes articles from }\textit{The Times of India}{~and }\textit{The
Indian Express}{. Further research can be conducted on the articles
posted by non--English language news media outlets, such as }\textit{Dainik
Bhaskar}\textit{~(Hindi language) and }\textit{Malayala Manorama}{~(Malayalam
language), which have the highest circulation in the country. Moreover,
researchers can find the most optimal keywords and filtering methods to
filter out articles that do not directly relate to the COVID--19
pandemic. For instance, our dataset included articles about Indian
elections that only mentioned transgender people as voter figures and
had no further mention of other LGBQ+ communities. Finally, researchers
can analyse comments present under the online editions of the articles
published by news media outlets to assess reader sentiment around the
articles. }

\bibliography{aaai22}

\begin{thebibliography}{70}
\providecommand{\natexlab}[1]{#1}

\bibitem[{Adamson et~al.(2022)Adamson, Lett, Glick, Garrison-Desany, and Restar}]{adamson2022experiences}
Adamson, T.; Lett, E.; Glick, J.; Garrison-Desany, H.~M.; and Restar, A. 2022.
\newblock Experiences of violence and discrimination among LGBTQ+ individuals during the COVID-19 pandemic: a global cross-sectional analysis.
\newblock \emph{BMJ global health}, 7(9): e009400.

\bibitem[{{\AA}kerlund(2019)}]{aakerlund2019representations}
{\AA}kerlund, M. 2019.
\newblock Representations of trans people in Swedish newspapers.
\newblock \emph{Journalism studies}, 20(9): 1319--1338.

\bibitem[{Al-Rawi et~al.(2021)Al-Rawi, Grepin, Li, Morgan, Wenham, and Smith}]{al2021investigating}
Al-Rawi, A.; Grepin, K.; Li, X.; Morgan, R.; Wenham, C.; and Smith, J. 2021.
\newblock Investigating public discourses around gender and COVID-19: a social media analysis of Twitter data.
\newblock \emph{Journal of Healthcare Informatics Research}, 5: 249--269.

\bibitem[{Alon et~al.(2020)Alon, Doepke, Olmstead-Rumsey, and Tertilt}]{alon2020impact}
Alon, T.; Doepke, M.; Olmstead-Rumsey, J.; and Tertilt, M. 2020.
\newblock The impact of COVID-19 on gender equality.
\newblock Technical report, National Bureau of economic research.

\bibitem[{Aneez et~al.(2019)Aneez, T~Neyazi, Kalogeropoulos, and Nielsen}]{aneez2019india}
Aneez, Z.; T~Neyazi, A.; Kalogeropoulos, A.; and Nielsen, R. 2019.
\newblock India digital news report.
\newblock \emph{Reuters}.

\bibitem[{Arakelyan, Arora, and Augenstein(2023)}]{arakelyan2023topic}
Arakelyan, E.; Arora, A.; and Augenstein, I. 2023.
\newblock Topic-Guided Sampling For Data-Efficient Multi-Domain Stance Detection.
\newblock \emph{arXiv preprint arXiv:2306.00765}.

\bibitem[{Arora and Kumar(2023)}]{arora2023media}
Arora, S.; and Kumar, K.~J. 2023.
\newblock \emph{Media Narratives and the COVID-19 Pandemic: The Asian Experience}.
\newblock Taylor \& Francis.

\bibitem[{{Audit Bureau of Circulations}(2023)}]{audit}
{Audit Bureau of Circulations}. 2023.
\newblock Highest Circulated Daily Newspapers (Languages Wise).
\newblock \url{http://www.auditbureau.org/files/JD%202022%20Highest%20Circulated%20(language%20wise).pdf}.
\newblock Accessed: 2024-05-13.

\bibitem[{Barker-Plummer(2013)}]{barker2013fixing}
Barker-Plummer, B. 2013.
\newblock Fixing Gwen: News and the mediation of (trans) gender challenges.
\newblock \emph{Feminist Media Studies}, 13(4): 710--724.

\bibitem[{Bogovi{\'c} et~al.(2021)Bogovi{\'c}, Me{\v{s}}trovi{\'c}, Beliga, and Martin{\v{c}}i{\'c}-Ip{\v{s}}i{\'c}}]{bogovic2021topic}
Bogovi{\'c}, P.~K.; Me{\v{s}}trovi{\'c}, A.; Beliga, S.; and Martin{\v{c}}i{\'c}-Ip{\v{s}}i{\'c}, S. 2021.
\newblock Topic modelling of Croatian news during COVID-19 pandemic.
\newblock In \emph{2021 44th International Convention on Information, Communication and Electronic Technology (MIPRO)}, 1044--1051. IEEE.

\bibitem[{Brown et~al.(2020)Brown, Mann, Ryder, Subbiah, Kaplan, Dhariwal, Neelakantan, Shyam, Sastry, Askell, Agarwal, Herbert-Voss, Krueger, Henighan, Child, Ramesh, Ziegler, Wu, Winter, Hesse, Chen, Sigler, Litwin, Gray, Chess, Clark, Berner, McCandlish, Radford, Sutskever, and Amodei}]{NEURIPS2020_1457c0d6}
Brown, T.; Mann, B.; Ryder, N.; Subbiah, M.; Kaplan, J.~D.; Dhariwal, P.; Neelakantan, A.; Shyam, P.; Sastry, G.; Askell, A.; Agarwal, S.; Herbert-Voss, A.; Krueger, G.; Henighan, T.; Child, R.; Ramesh, A.; Ziegler, D.; Wu, J.; Winter, C.; Hesse, C.; Chen, M.; Sigler, E.; Litwin, M.; Gray, S.; Chess, B.; Clark, J.; Berner, C.; McCandlish, S.; Radford, A.; Sutskever, I.; and Amodei, D. 2020.
\newblock Language Models are Few-Shot Learners.
\newblock In Larochelle, H.; Ranzato, M.; Hadsell, R.; Balcan, M.; and Lin, H., eds., \emph{Advances in Neural Information Processing Systems}, volume~33, 1877--1901. Curran Associates, Inc.

\bibitem[{Capuzza(2016)}]{capuzza2016improvements}
Capuzza, J.~C. 2016.
\newblock Improvements still needed for transgender coverage.
\newblock \emph{Newspaper Research Journal}, 37(1): 82--94.

\bibitem[{Carli(2020)}]{carli2020women}
Carli, L.~L. 2020.
\newblock Women, gender equality and COVID-19.
\newblock \emph{Gender in management: an International Journal}, 35(7/8): 647--655.

\bibitem[{Chakraborty et~al.(2016)Chakraborty, Paranjape, Kakarla, and Ganguly}]{chakraborty2016stop}
Chakraborty, A.; Paranjape, B.; Kakarla, S.; and Ganguly, N. 2016.
\newblock Stop clickbait: Detecting and preventing clickbaits in online news media.
\newblock In \emph{2016 IEEE/ACM international conference on advances in social networks analysis and mining (ASONAM)}, 9--16. IEEE.

\bibitem[{Fairclough(2013)}]{fairclough2013language}
Fairclough, N. 2013.
\newblock \emph{Language and power}.
\newblock Routledge.

\bibitem[{Fish et~al.(2021)Fish, Salerno, Williams, Rinderknecht, Drotning, Sayer, and Doan}]{fish2021sexual}
Fish, J.~N.; Salerno, J.; Williams, N.~D.; Rinderknecht, R.~G.; Drotning, K.~J.; Sayer, L.; and Doan, L. 2021.
\newblock Sexual minority disparities in health and well-being as a consequence of the COVID-19 pandemic differ by sexual identity.
\newblock \emph{LGBT health}, 8(4): 263--272.

\bibitem[{Flor et~al.(2022)Flor, Friedman, Spencer, Cagney, Arrieta, Herbert, Stein, Mullany, Hon, Patwardhan et~al.}]{flor2022quantifying}
Flor, L.~S.; Friedman, J.; Spencer, C.~N.; Cagney, J.; Arrieta, A.; Herbert, M.~E.; Stein, C.; Mullany, E.~C.; Hon, J.; Patwardhan, V.; et~al. 2022.
\newblock Quantifying the effects of the COVID-19 pandemic on gender equality on health, social, and economic indicators: a comprehensive review of data from March, 2020, to September, 2021.
\newblock \emph{The Lancet}, 399(10344): 2381--2397.

\bibitem[{Ganguly and Singh(2021)}]{ganguly2021transgender}
Ganguly, D.; and Singh, R. 2021.
\newblock The transgender humanitarian crisis during the Covid-19 pandemic in India.
\newblock \emph{Intersections: Gender and Sexuality in Asia and the Pacific}.

\bibitem[{Gausman and Langer(2020)}]{gausman2020sex}
Gausman, J.; and Langer, A. 2020.
\newblock Sex and gender disparities in the COVID-19 pandemic.
\newblock \emph{Journal of Women's Health}, 29(4): 465--466.

\bibitem[{Ghasiya and Okamura(2021)}]{ghasiya2021investigating}
Ghasiya, P.; and Okamura, K. 2021.
\newblock Investigating COVID-19 news across four nations: A topic modeling and sentiment analysis approach.
\newblock \emph{Ieee Access}, 9: 36645--36656.

\bibitem[{Grootendorst(2022)}]{grootendorst2022bertopic}
Grootendorst, M. 2022.
\newblock BERTopic: Neural topic modeling with a class-based TF-IDF procedure.
\newblock \emph{arXiv preprint arXiv:2203.05794}.

\bibitem[{Gupta(2019)}]{gupta2019response}
Gupta, K. 2019.
\newblock Response and responsibility: Mainstream media and Lucy Meadows in a post-Leveson context.
\newblock \emph{Sexualities}, 22(1-2): 31--47.

\bibitem[{Hackl, Becker, and Todd(2016)}]{hackl2016chelsea}
Hackl, A.~M.; Becker, A.~B.; and Todd, M.~E. 2016.
\newblock “I am Chelsea Manning”: Comparison of gendered representation of Private Manning in US and international news media.
\newblock \emph{Journal of homosexuality}, 63(4): 467--486.

\bibitem[{Hardalov et~al.(2021)Hardalov, Arora, Nakov, and Augenstein}]{hardalov2021cross}
Hardalov, M.; Arora, A.; Nakov, P.; and Augenstein, I. 2021.
\newblock Cross-domain label-adaptive stance detection.
\newblock \emph{arXiv preprint arXiv:2104.07467}.

\bibitem[{Hindarto(2022)}]{hindarto2022investigating}
Hindarto, I.~H. 2022.
\newblock Investigating How the National Online Media Reported the LGBT Community During the COVID-19 Pandemic.
\newblock \emph{Profetik: Jurnal Komunikasi}, 15(2): 208--227.

\bibitem[{Hiscott et~al.(2020)Hiscott, Alexandridi, Muscolini, Tassone, Palermo, Soultsioti, and Zevini}]{hiscott2020global}
Hiscott, J.; Alexandridi, M.; Muscolini, M.; Tassone, E.; Palermo, E.; Soultsioti, M.; and Zevini, A. 2020.
\newblock The global impact of the coronavirus pandemic.
\newblock \emph{Cytokine \& growth factor reviews}, 53: 1--9.

\bibitem[{{Hu, Krystal}(2023)}]{hu}
{Hu, Krystal}. 2023.
\newblock ChatGPT Sets Record for Fastest-Growing User Base - Analyst Note.
\newblock \url{http://www.reuters.com/technology/chatgpt-sets-record-fastest-growing-user-base-analyst-note-2023-02-01/}.
\newblock Accessed: 2024-05-13.

\bibitem[{Jain et~al.(2021)Jain, Dey, Kelkar, and Ahlawat}]{jain2021analysis}
Jain, J.; Dey, D.; Kelkar, B.; and Ahlawat, K. 2021.
\newblock Analysis of Indian News with Corona Headlines Classification.
\newblock In \emph{International Conference on Artificial Intelligence and Speech Technology}, 116--126. Springer.

\bibitem[{Jiang et~al.(2019)Jiang, Guo, Xu, Zhao, and Fu}]{jiang2019prompts}
Jiang, T.; Guo, Q.; Xu, Y.; Zhao, Y.; and Fu, S. 2019.
\newblock What prompts users to click on news headlines? A clickstream data analysis of the effects of news recency and popularity.
\newblock In \emph{International Conference on Information}, 539--546. Springer.

\bibitem[{Koco{\'n} et~al.(2023)Koco{\'n}, Cichecki, Kaszyca, Kochanek, Szyd{\l}o, Baran, Bielaniewicz, Gruza, Janz, Kanclerz et~al.}]{kocon2023chatgpt}
Koco{\'n}, J.; Cichecki, I.; Kaszyca, O.; Kochanek, M.; Szyd{\l}o, D.; Baran, J.; Bielaniewicz, J.; Gruza, M.; Janz, A.; Kanclerz, K.; et~al. 2023.
\newblock ChatGPT: Jack of all trades, master of none.
\newblock \emph{Information Fusion}, 99: 101861.

\bibitem[{Konnoth(2020)}]{konnoth2020supporting}
Konnoth, C. 2020.
\newblock Supporting LGBT communities in the COVID-19 pandemic.
\newblock \emph{2020). Assessing Legal Responses to COVID-19. Boston: Public Health Law Watch, U of Colorado Law Legal Studies Research Paper}, 20--47.

\bibitem[{{KPMG India}(2020)}]{kpmg2020media}
{KPMG India}. 2020.
\newblock Synopsis: KPMG India Media and Entertainment Report 2020.
\newblock \url{https://assets.kpmg.com/content/dam/kpmg/in/pdf/2020/09/synopsis-kpmg-india-media-and-entertainment-2020.pdf}.
\newblock Accessed: 2024-08-25.

\bibitem[{K{\"u}{\c{c}}{\"u}k and Can(2020)}]{kuccuk2020stance}
K{\"u}{\c{c}}{\"u}k, D.; and Can, F. 2020.
\newblock Stance detection: A survey.
\newblock \emph{ACM Computing Surveys (CSUR)}, 53(1): 1--37.

\bibitem[{Kumaresan et~al.(2023)Kumaresan, Ponnusamy, Priyadharshini, Buitelaar, and Chakravarthi}]{kumaresan2023homophobia}
Kumaresan, P.~K.; Ponnusamy, R.; Priyadharshini, R.; Buitelaar, P.; and Chakravarthi, B.~R. 2023.
\newblock Homophobia and transphobia detection for low-resourced languages in social media comments.
\newblock \emph{Natural Language Processing Journal}, 5: 100041.

\bibitem[{Listiorini and Vidiadari(2022)}]{listiorini2022news}
Listiorini, D.; and Vidiadari, I.~S. 2022.
\newblock News of LGBT on online media in 2020: endless stigma.
\newblock \emph{Jurnal Studi Komunikasi}, 6(2): 531--546.

\bibitem[{Liu et~al.(2020)Liu, Zheng, Zheng, Chen, Liu, Chen, Chu, Zhu, Akinwunmi, Huang et~al.}]{liu2020health}
Liu, Q.; Zheng, Z.; Zheng, J.; Chen, Q.; Liu, G.; Chen, S.; Chu, B.; Zhu, H.; Akinwunmi, B.; Huang, J.; et~al. 2020.
\newblock Health communication through news media during the early stage of the COVID-19 outbreak in China: digital topic modeling approach.
\newblock \emph{Journal of medical Internet research}, 22(4): e19118.

\bibitem[{Liu et~al.(2019)Liu, Ott, Goyal, Du, Joshi, Chen, Levy, Lewis, Zettlemoyer, and Stoyanov}]{liu2019roberta}
Liu, Y.; Ott, M.; Goyal, N.; Du, J.; Joshi, M.; Chen, D.; Levy, O.; Lewis, M.; Zettlemoyer, L.; and Stoyanov, V. 2019.
\newblock Roberta: A robustly optimized bert pretraining approach.
\newblock \emph{arXiv preprint arXiv:1907.11692}.

\bibitem[{Lucas et~al.(2022)Lucas, Bouchoucha, Afrouz, Reed, and Brennan-Olsen}]{lucas2022lgbtq+}
Lucas, J.~J.; Bouchoucha, S.~L.; Afrouz, R.; Reed, K.; and Brennan-Olsen, S.~L. 2022.
\newblock LGBTQ+ loss and grief in a cis-heteronormative pandemic: a qualitative evidence synthesis of the COVID-19 literature.
\newblock \emph{Qualitative Health Research}, 32(14): 2102--2117.

\bibitem[{MacKenzie and Marcel(2009)}]{mackenzie2009media}
MacKenzie, G.; and Marcel, M. 2009.
\newblock Media coverage of the murder of US transwomen of color.
\newblock \emph{Local violence, global media: Feminist analyses of gendered representations}, 79--106.

\bibitem[{McQuail(1987)}]{mcquail1987mass}
McQuail, D. 1987.
\newblock \emph{Mass communication theory: An introduction}.
\newblock Sage Publications, Inc.

\bibitem[{{Media Research Users Council}(2019)}]{irs2019}
{Media Research Users Council}, R. 2019.
\newblock Indian Readership Survey Q4 2019 Highlights.
\newblock Accessed: 2024-09-05.

\bibitem[{Mohammad et~al.(2016)Mohammad, Kiritchenko, Sobhani, Zhu, and Cherry}]{mohammad2016semeval}
Mohammad, S.; Kiritchenko, S.; Sobhani, P.; Zhu, X.; and Cherry, C. 2016.
\newblock Semeval-2016 task 6: Detecting stance in tweets.
\newblock In \emph{Proceedings of the 10th international workshop on semantic evaluation (SemEval-2016)}, 31--41.

\bibitem[{Morrison et~al.(2021)Morrison, Parker, Sadika, Sameen, and Morrison}]{morrison2021newsworthy}
Morrison, M.~A.; Parker, K.~M.; Sadika, B.; Sameen, D.-E.; and Morrison, T.~G. 2021.
\newblock ‘Newsworthy enough?’: media framing of Canadian LGBTQ persons’ sexual violence experiences.
\newblock \emph{Psychology \& Sexuality}, 12(1-2): 96--114.

\bibitem[{Newman et~al.(2021)Newman, Fletcher, Schulz, Andi, Robertson, and Nielsen}]{newman2021reuters}
Newman, N.; Fletcher, R.; Schulz, A.; Andi, S.; Robertson, C.~T.; and Nielsen, R.~K. 2021.
\newblock Reuters Institute digital news report 2021.
\newblock \emph{Reuters Institute for the study of Journalism}.

\bibitem[{Organization et~al.(2022)}]{world2022director}
Organization, W.~H.; et~al. 2022.
\newblock WHO Director-General’s opening remarks at the media briefing on COVID-19.
\newblock \emph{January}, 30.

\bibitem[{Palau~Sampio(2016)}]{palau2016reference}
Palau~Sampio, D. 2016.
\newblock Reference press metamorphosis in the digital context: clickbait and tabloid strategies in Elpais. com.
\newblock \emph{Communication \& Society}, 29(2).

\bibitem[{Phillips~Ii et~al.(2020)Phillips~Ii, Felt, Ruprecht, Wang, Xu, P{\'e}rez-Bill, Bagnarol, Roth, Curry, and Beach}]{phillips2020addressing}
Phillips~Ii, G.; Felt, D.; Ruprecht, M.~M.; Wang, X.; Xu, J.; P{\'e}rez-Bill, E.; Bagnarol, R.~M.; Roth, J.; Curry, C.~W.; and Beach, L.~B. 2020.
\newblock Addressing the disproportionate impacts of the COVID-19 pandemic on sexual and gender minority populations in the United States: actions toward equity.
\newblock \emph{LGBT health}, 7(6): 279--282.

\bibitem[{Pomerleau and Rao(2017)}]{pomerleau2017fake}
Pomerleau, D.; and Rao, D. 2017.
\newblock Fake news challenge stage 1 (FNC-I): Stance detection.
\newblock \emph{URL www. fakenewschallenge. org}.

\bibitem[{Qin et~al.(2023)Qin, Zhang, Zhang, Chen, Yasunaga, and Yang}]{qin2023chatgpt}
Qin, C.; Zhang, A.; Zhang, Z.; Chen, J.; Yasunaga, M.; and Yang, D. 2023.
\newblock Is chatgpt a general-purpose natural language processing task solver?
\newblock \emph{arXiv preprint arXiv:2302.06476}.

\bibitem[{Raj(2023)}]{raj}
Raj, S. 2023.
\newblock Gender, Media and the Covid-19 Pandemic.
\newblock \emph{Media Narratives and the COVID-19 Pandemic: The Asian Experience}.

\bibitem[{Rieis et~al.(2015)Rieis, de~Souza, de~Melo, Prates, Kwak, and An}]{rieis2015breaking}
Rieis, J.; de~Souza, F.; de~Melo, P.~V.; Prates, R.; Kwak, H.; and An, J. 2015.
\newblock Breaking the news: First impressions matter on online news.
\newblock In \emph{Proceedings of the international AAAI conference on web and social media}, 357--366.

\bibitem[{{Roy, Raina}(2020)}]{roy}
{Roy, Raina}. 2020.
\newblock Coronavirus: Kolkata’s Trans Community Has Been Locked out of Healthcare and Livelihood.
\newblock \url{https://scroll.in/article/968182/coronavirus-kolkatas-trans-community-has-been-locked-out-of-healthcare-and-livelihood}.
\newblock Accessed: 2024-05-14.

\bibitem[{Saxena(2021)}]{saxena2021impact}
Saxena, V.~K. 2021.
\newblock Impact of Covid-19 on Indian national print media.
\newblock \emph{International Journal of Research and Analytical Reviews (IJRAR)}.

\bibitem[{Scacco and Muddiman(2016)}]{scacco2016investigating}
Scacco, J.~M.; and Muddiman, A. 2016.
\newblock Investigating the influence of “clickbait” news headlines.
\newblock \emph{Engaging News Project Report}.

\bibitem[{Semykina(2018)}]{semykina2018media}
Semykina, K. 2018.
\newblock Media Construction of LGBT Prides in Russia: Framing Dynamics and Frame Resonance.
\newblock \emph{Higher School of Economics Research Paper No. WP BRP}, 81.

\bibitem[{Shah et~al.(2015)Shah, Merchant, Mahajan, and Nevatia}]{shah2015no}
Shah, C.; Merchant, R.; Mahajan, S.; and Nevatia, S. 2015.
\newblock \emph{No outlaws in the gender galaxy}.
\newblock Zubaan.

\bibitem[{Stefanov et~al.(2020)Stefanov, Darwish, Atanasov, and Nakov}]{stefanov2020predicting}
Stefanov, P.; Darwish, K.; Atanasov, A.; and Nakov, P. 2020.
\newblock Predicting the topical stance and political leaning of media using tweets.
\newblock In \emph{Proceedings of the 58th Annual Meeting of the Association for Computational Linguistics}, 527--537.

\bibitem[{Stemple, Karegeya, and Gruskin(2016)}]{stemple2016human}
Stemple, L.; Karegeya, P.; and Gruskin, S. 2016.
\newblock Human rights, gender, and infectious disease: from HIV/AIDS to Ebola.
\newblock \emph{Hum. Rts. Q.}, 38: 993.

\bibitem[{Stevens, Acic, and Rhea(2021)}]{stevens2021natural}
Stevens, H.~R.; Acic, I.; and Rhea, S. 2021.
\newblock Natural language processing insight into LGBTQ+ youth mental health during the COVID-19 pandemic: longitudinal content analysis of anxiety-provoking topics and trends in emotion in LGBTeens microcommunity subreddit.
\newblock \emph{JMIR public health and surveillance}, 7(8): e29029.

\bibitem[{{The Times of India}(2023)}]{times}
{The Times of India}. 2023.
\newblock Times Out \& Proud: It’s time to claim your space with pride.
\newblock \url{https://timesofindia.indiatimes.com/Times-Out-Proud-Its-time-to-claim-your-space-with-pride/campaignlanding/69161947.cms?from=mdr/}.
\newblock Accessed: 2024-05-14.

\bibitem[{Tshuma, Tshuma, and Ndlovu(2022)}]{tshuma2022media}
Tshuma, B.~B.; Tshuma, L.~A.; and Ndlovu, N. 2022.
\newblock Media discourses on gender in the time of COVID-19 pandemic in Zimbabwe.
\newblock \emph{Health crises and media discourses in Sub-Saharan Africa}, 267.

\bibitem[{{Velasco, Gabi and Langness, Mel}(2020)}]{langness}
{Velasco, Gabi and Langness, Mel}. 2020.
\newblock COVID-19 Action That Centers Black LGBTQ People Can Address Housing Inequities.
\newblock \url{https://www.urban.org/urban-wire/covid-19-action-centers-black-lgbtq-people-can-address-housing-inequities}.
\newblock Accessed: 2024-05-14.

\bibitem[{Wan et~al.(2021)Wan, Lucic, Ghazzai, and Massoud}]{wan2021topic}
Wan, X.; Lucic, M.~C.; Ghazzai, H.; and Massoud, Y. 2021.
\newblock Topic modeling and progression of American digital news media during the onset of the COVID-19 pandemic.
\newblock \emph{IEEE Transactions on Technology and Society}, 3(2): 111--120.

\bibitem[{Wang et~al.(2023)Wang, Xie, Feng, Ding, Yang, and Xia}]{wang2023chatgpt}
Wang, Z.; Xie, Q.; Feng, Y.; Ding, Z.; Yang, Z.; and Xia, R. 2023.
\newblock Is ChatGPT a good sentiment analyzer? A preliminary study.
\newblock \emph{arXiv preprint arXiv:2304.04339}.

\bibitem[{Wenham et~al.(2020)Wenham, Smith, Davies, Feng, Gr{\'e}pin, Harman, Herten-Crabb, and Morgan}]{wenham2020women}
Wenham, C.; Smith, J.; Davies, S.~E.; Feng, H.; Gr{\'e}pin, K.~A.; Harman, S.; Herten-Crabb, A.; and Morgan, R. 2020.
\newblock Women are most affected by pandemics—lessons from past outbreaks.
\newblock \emph{Nature}, 583(7815): 194--198.

\bibitem[{Whittington, Hadfield, and Calder{\'o}n(2020)}]{whittington2020lives}
Whittington, C.; Hadfield, K.; and Calder{\'o}n, C. 2020.
\newblock \emph{The lives \& livelihoods of many in the LGBTQ community are at risk amidst COVID-19 crisis}.
\newblock Human Rights Campaign Foundation.

\bibitem[{Wiktorek(2015)}]{wiktorek2015prohibit}
Wiktorek, A.~E. 2015.
\newblock \emph{Prohibit, tolerate, or prefer: a content analysis of agenda-setting and the LGBT in MSNBC and Fox News}.
\newblock Liberty University.

\bibitem[{Willox(2003)}]{willox2003branding}
Willox, D. 2003.
\newblock Branding Teena:(Mis) representations in the media.
\newblock \emph{Sexualities}, 6(3-4): 407--425.

\bibitem[{{World Health Organization}(2023)}]{who23}
{World Health Organization}. 2023.
\newblock WHO COVID-19 dashboard.
\newblock \url{https://data.who.int/dashboards/covid19/vaccines?n=c}.
\newblock Accessed: 2024-01-13.

\bibitem[{Yuan et~al.(2023)Yuan, Verma, Keller, and Aledavood}]{yuan2023minority}
Yuan, Y.; Verma, G.; Keller, B.; and Aledavood, T. 2023.
\newblock Minority stress experienced by LGBTQ online communities during the COVID-19 pandemic.
\newblock In \emph{Proceedings of the International AAAI Conference on Web and Social Media}, volume~17, 936--947.

\end{thebibliography}


\subsection{Ethics Checklist}

\begin{enumerate}

\item For most authors...
\begin{enumerate}
    \item  Would answering this research question advance science without violating social contracts, such as violating privacy norms, perpetuating unfair profiling, exacerbating the socio-economic divide, or implying disrespect to societies or cultures?
    \answerYes{Yes}
  \item Do your main claims in the abstract and introduction accurately reflect the paper's contributions and scope?
    \answerYes{Yes}
   \item Do you clarify how the proposed methodological approach is appropriate for the claims made? 
    \answerYes{Yes}
   \item Do you clarify what are possible artifacts in the data used, given population-specific distributions?
    \answerNA{No, data was obtained from publicly available news archives.}
  \item Did you describe the limitations of your work?
    \answerYes{Yes}
  \item Did you discuss any potential negative societal impacts of your work?
    \answerNA{NA}
      \item Did you discuss any potential misuse of your work?
    \answerNA{NA}
    \item Did you describe steps taken to prevent or mitigate potential negative outcomes of the research, such as data and model documentation, data anonymization, responsible release, access control, and the reproducibility of findings?
    \answerYes{Yes}
  \item Have you read the ethics review guidelines and ensured that your paper conforms to them?
    \answerYes{Yes}
\end{enumerate}

\item Additionally, if your study involves hypotheses testing...
\begin{enumerate}
  \item Did you clearly state the assumptions underlying all theoretical results?
    \answerYes{Yes}
  \item Have you provided justifications for all theoretical results?
    \answerYes{Yes}
  \item Did you discuss competing hypotheses or theories that might challenge or complement your theoretical results?
    \answerNA{NA}
  \item Have you considered alternative mechanisms or explanations that might account for the same outcomes observed in your study?
    \answerYes{Yes}
  \item Did you address potential biases or limitations in your theoretical framework?
    \answerYes{Yes}
  \item Have you related your theoretical results to the existing literature in social science?
    \answerYes{Yes}
  \item Did you discuss the implications of your theoretical results for policy, practice, or further research in the social science domain?
    \answerYes{Yes}
\end{enumerate}

\item Additionally, if you are including theoretical proofs...
\begin{enumerate}
  \item Did you state the full set of assumptions of all theoretical results?
    \answerNA{NA}
	\item Did you include complete proofs of all theoretical results?
    \answerNA{NA}
\end{enumerate}

\item Additionally, if you ran machine learning experiments...
\begin{enumerate}
  \item Did you include the code, data, and instructions needed to reproduce the main experimental results (either in the supplemental material or as a URL)?
    \answerYes{Yes}
  \item Did you specify all the training details (e.g., data splits, hyperparameters, how they were chosen)?
    \answerYes{Yes}
     \item Did you report error bars (e.g., with respect to the random seed after running experiments multiple times)?
    \answerYes{Yes}
	\item Did you include the total amount of compute and the type of resources used (e.g., type of GPUs, internal cluster, or cloud provider)?
    \answerNA{No, personal computers were used for running the experiments}
     \item Do you justify how the proposed evaluation is sufficient and appropriate to the claims made? 
    \answerYes{Yes}
     \item Do you discuss what is ``the cost`` of misclassification and fault (in)tolerance?
    \answerYes{Yes}
  
\end{enumerate}

\item Additionally, if you are using existing assets (e.g., code, data, models) or curating/releasing new assets, \textbf{without compromising anonymity}...
\begin{enumerate}
  \item If your work uses existing assets, did you cite the creators?
    \answerYes{Yes}
  \item Did you mention the license of the assets?
    \answerNo{No}
  \item Did you include any new assets in the supplemental material or as a URL?
    \answerYes{Yes}
  \item Did you discuss whether and how consent was obtained from people whose data you're using/curating?
    \answerNA{Data from newspaper websites were retrieved in accordance to the robots.txt files.}
  \item Did you discuss whether the data you are using/curating contains personally identifiable information or offensive content?
    \answerNA{Data was obtained from news archives from which author information etc was not retrieved}
\item If you are curating or releasing new datasets, did you discuss how you intend to make your datasets FAIR?
\answerNA{NA}
\item If you are curating or releasing new datasets, did you create a Datasheet for the Dataset? 
\answerNA{NA}
\end{enumerate}

\item Additionally, if you used crowdsourcing or conducted research with human subjects, \textbf{without compromising anonymity}...
\begin{enumerate}
  \item Did you include the full text of instructions given to participants and screenshots?
    \answerNA{No, annotators were not given instructions to avoid data tampering}
  \item Did you describe any potential participant risks, with mentions of Institutional Review Board (IRB) approvals?
    \answerNA{NA}
  \item Did you include the estimated hourly wage paid to participants and the total amount spent on participant compensation?
    \answerNA{NA}
   \item Did you discuss how data is stored, shared, and deidentified?
   \answerNA{The annotators were only given the article headline for the task of sentiment analysis}
\end{enumerate}

\end{enumerate}
\section{Appendix}
\label{sec:appendix}

\appendix
\section{Challenges using ChatGPT}
\label{sec:appendixGPT}

{For various reasons, it took us multiple prompts to get the desired
outputs.}{~}{First, ChatGPT used commas as the delimiter for the output,
which led to multiple columns for headlines with more than one comma.
}{Additionally, ChatGPT kept forgetting the task and would randomly
summarise the headlines and output sentiments of any randomly chosen ten
headlines instead of performing sentiment analysis. Furthermore, it
refused to provide sentiment for a headline without further
context.} {ChatGPT\textquotesingle s
outputs may vary with every launch, even with the same prompt \cite{kocon2023chatgpt}. All outputs from the rephrased instruction were considered
the final output from ChatGPT.}{~For the next step, we made sure all of
the headlines given to ChatGPT were given an output. }{The ones that
ChatGPT skipped in the process were sent again in the same
conversation.}
{Link to the conversation: {\textcolor[HTML]{000099} {https://shorturl.at/fhoyW}}}

\section{Class--wise metrics}
\label{sec:appendixCM}

Tables \ref{table5} and \ref{table6} represent the class--wise metrics of distil--RoBERTa--base and ChatGPT–3.5.

\begin{table}[htbp]
\centering
\resizebox{.95\columnwidth}{!}{

\begin{tabular}{|c|c|c|c|}
\hline
\textbf{Sentiment} & \textbf{F1-score} & \textbf{Precision} & \textbf{Recall} \\
\hline
Positive & 0.77 & 0.72 & 0.83 \\
Neutral & 0.79 & 0.92 & 0.69 \\
Negative & 0.76 & 0.64 & 0.93 \\
\hline
\end{tabular}
}
\caption{Class–wise metrics for evaluation of results from distil-RoBERTa-base}
\label{table5}
\end{table}

\begin{table}[htbp]
\centering
\resizebox{.95\columnwidth}{!}{

\begin{tabular}{|c|c|c|c|}
\hline
\textbf{Sentiment} & \textbf{F1-score} & \textbf{Precision} & \textbf{Recall} \\
\hline
Positive & 0.72 & 0.64 & 0.83 \\
Neutral & 0.77 & 0.80 & 0.74 \\
Negative & 0.75 & 0.83 & 0.68 \\
\hline
\end{tabular}
}
\caption{Class–wise metrics for evaluation of results from ChatGPT}
\label{table6}
\end{table}

\section{Sentiment Analysis}
\label{sec:appendixSA}
\noindent Tables \ref{table17} and \ref{table18} re
present the polarity of the content of the articles assigned by distil--RoBERTa--base to articles published by \textit{TOI} and \textit{IE} both during and before the pandemic. 

\begin{table}[htbp]
\centering
\resizebox{.95\columnwidth}{!}{

\begin{tabular}{|c|c|c|}
\hline
\textbf{Sentiment} & \textbf{TOI} & \textbf{TOI} \\
\textbf{(in \%)} & \textbf{[during pandemic]} & \textbf{[before pandemic]} \\
\hline
Positive & 41.1 & 38.0\\
Negative & 31.6 & 27.7 \\
Neutral & 27.3 & 34.3 \\
\hline
\end{tabular}
}
\caption{The polarity of content of \textit{TOI} news articles assigned by distil–RoBERTa–base}
\label{table17}
\end{table}

\begin{table}[pb]
\centering
\resizebox{.95\columnwidth}{!}{

\begin{tabular}{|c|c|c|}
\hline
\textbf{Sentiment} & \textbf{IE} & \textbf{IE} \\
\textbf{(in \%)} & \textbf{[during pandemic]} & \textbf{[before pandemic]} \\
\hline
Positive  & 33.2 & 28.0  \\
Negative & 31.2 & 33.5 \\
Neutral & 35.6 & 38.5 \\
\hline
\end{tabular}
}
\caption{The polarity of content of \textit{IE} news articles assigned by distil–RoBERTa–base}
\label{table18}
\end{table}
\end{document}